\documentclass[%
    a4paper,fleqn,  
    twocolumn,
]{article}

\usepackage{amsmath}

\usepackage{txfonts}%% for arXiv

\usepackage{graphicx}
\usepackage{siunitx}
\usepackage{cite}
\usepackage[setpagesize=false,
]{hyperref}

\usepackage{pfr}

\let\bm\boldsymbol
\usepackage{xparse}
\providecommand\difsymbol{\mathrm{d}}
\providecommand\Difsymbol{\mathrm{D}}
\providecommand\pdifsymbol{\partial}
\providecommand\vdifsymbol{\delta}
\providecommand\innerfrac[2]{#1/#2}
\NewDocumentCommand\dd{o g}{
    \IfNoValueTF{#2}{
        \difsymbol\IfNoValueTF{#1}{}{^{#1}}
    }{
        \mathinner{\difsymbol\IfNoValueTF{#1}{}{^{#1}}#2}
    }
}
\NewDocumentCommand\DD{o g}{
    \IfNoValueTF{#2}{
        \Difsymbol\IfNoValueTF{#1}{}{^{#1}}
    }{
        \mathinner{\Difsymbol\IfNoValueTF{#1}{}{^{#1}}#2}
    }
}
\NewDocumentCommand\pd{o g}{
    \IfNoValueTF{#2}{
        \pdifsymbol\IfNoValueTF{#1}{}{^{#1}}
    }{
        \mathinner{\pdifsymbol\IfNoValueTF{#1}{}{^{#1}}#2}
    }
}
\NewDocumentCommand\vd{o g}{
    \IfNoValueTF{#2}{
        \vdifsymbol\IfNoValueTF{#1}{}{^{#1}}
    }{
        \mathinner{\vdifsymbol\IfNoValueTF{#1}{}{^{#1}}#2}
    }
}
\NewDocumentCommand\dif{s o m g g}{
    \IfBooleanTF{#1}{\let\fractype\innerfrac}{\let\fractype\frac}
    \IfNoValueTF{#4}{
        \fractype{\difsymbol\IfNoValueTF{#2}{}{^{#2}}}{\difsymbol#3\IfNoValueTF{#2}{}{^{#2}}}
    }{
        \IfNoValueTF{#5}{
            \fractype{\difsymbol\IfNoValueTF{#2}{}{^{#2}}#3}{\difsymbol#4\IfNoValueTF{#2}{}{^{#2}}}
        }{
            \fractype{\difsymbol^2#3}{\difsymbol#4\difsymbol#5}
        }
    }
}
\NewDocumentCommand\pdif{s o m g g}{
    \IfBooleanTF{#1}{\let\fractype\innerfrac}{\let\fractype\frac}
    \IfNoValueTF{#4}{
        \fractype{\pdifsymbol\IfNoValueTF{#2}{}{^{#2}}}{\pdifsymbol#3\IfNoValueTF{#2}{}{^{#2}}}
    }{
        \IfNoValueTF{#5}{
            \fractype{\pdifsymbol\IfNoValueTF{#2}{}{^{#2}}#3}{\pdifsymbol#4\IfNoValueTF{#2}{}{^{#2}}}
        }{
            \fractype{\pdifsymbol^2#3}{\pdifsymbol#4\pdifsymbol#5}
        }
    }
}
\NewDocumentCommand\vdif{s o m g g}{
    \IfBooleanTF{#1}{\let\fractype\innerfrac}{\let\fractype\frac}
    \IfNoValueTF{#4}{
        \fractype{\vdifsymbol\IfNoValueTF{#2}{}{^{#2}}}{\vdifsymbol#3\IfNoValueTF{#2}{}{^{#2}}}
    }{
        \IfNoValueTF{#5}{
            \fractype{\vdifsymbol\IfNoValueTF{#2}{}{^{#2}}#3}{\vdifsymbol#4\IfNoValueTF{#2}{}{^{#2}}}
        }{
            \fractype{\vdifsymbol^2#3}{\vdifsymbol#4\vdifsymbol#5}
        }
    }
}
\NewDocumentCommand\Dif{s o m g g}{
    \IfBooleanTF{#1}{\let\fractype\innerfrac}{\let\fractype\frac}
    \IfNoValueTF{#4}{
        \fractype{\Difsymbol\IfNoValueTF{#2}{}{^{#2}}}{\Difsymbol#3\IfNoValueTF{#2}{}{^{#2}}}
    }{
        \IfNoValueTF{#5}{
            \fractype{\Difsymbol\IfNoValueTF{#2}{}{^{#2}}#3}{\Difsymbol#4\IfNoValueTF{#2}{}{^{#2}}}
        }{
            \fractype{\Difsymbol^2#3}{\vDfsymbol#4\Difsymbol#5}
        }
    }
}

\newcommand{\rme}{\mathrm{e}}

\newcommand{\rmv}{\mathrm{v}}
\newcommand{\rms}{\mathrm{s}}
\newcommand{\rmw}{\mathrm{w}}
\newcommand{\rmD}{\mathrm{D}}

\newcommand{\calM}{\mathcal{M}}

\begin{document}

\title{Enhanced entropy production in heat-flux-driven plasma sheath}

\author{Yuji OHNO and Zensho YOSHIDA}

%\affiliation{Graduate School of Frontier Sciences, The University of Tokyo, 5-1-5 Kashiwanoha, Kashiwa 277-8561, Japan}
\affiliation{Graduate School of Frontier Sciences, The University of Tokyo, 5-1-5 Kashiwanoha, Kashiwa, Chiba 277-8561, Japan}

\email{ohno@ppl.k.u-tokyo.ac.jp}

\date{(%
    %Received / Accepted %
    \today%
)}

\begin{abstract}
The plasma sheath sets a stage for 
a strongly nonlinear coupling of the thermal, kinetic, and electric energies of plasma in a non-equilibrium, open environment.
The pressure, velocity, and electrostatic potential profiles 
depend strongly on the boundary condition given on the internal side (pre-sheath).
By controlling the boundary values of the heat flux and the ion Mach number, we solve a set of equations for the ion temperature and the electrostatic potential. 
The boundary values of the ion velocity and the electrostatic potential vary due to the change of the boundary ion temperature. 
When the heat flux exceeds a threshold value (determined by the ion Mach number),
the temperature contrast is enhanced, resulting in a large entropy production. 
\end{abstract}


\DOI{10.1585/pfr.xx.xxxxx}

\maketitle  % Don't forget to put this!

%%%%% MAIN TEXT %%%%%

A non-equilibrium plasma structure called sheath forms 
when a plasma contacts with a material wall which absorbs charged particles \cite{Stangeby2000,Riemann:sheath-review}. 
Studies of non-equilibrium thermodynamics have focused on the entropy production (EP) as a determinant of self-organized structures (see, e.g., Refs.~\cite{Ozawa2001:PRE,ShimokawaOzawa2001,Niven2009:PRE,Dewar2014:EPbook}).
In plasma physics, works on bifurcations in thermodynamical models for self-organized structures
%, such as zonal flow and B\'enard convection, 
have been developed \cite{YoshidaMahajan2008:MEP,KawazuraEPR1,KawazuraEPR2,YoshidaKawazuraEPR}. 
Consequences of the works indicate that the bifurcation property (maximization or minimization of EP) 
changes depending on the driving condition---which of heat flux and temperature at the boundary is controlled. 
In a fusion device, hot plasmas bring large heat fluxes to a sheath. 
In this work, we study the response of a sheath to a heat transport.
We consider the thermal energy in addition to the kinetic and electric energies.
Moreover, we introduce an irreversible heat flux and solve a flux-driven system. 

For simplicity, we assume that 
the electron density $n_\rme$ obeys the Boltzmann distribution with a constant temperature $T_\rme$  ($n_\rme = n_0 \exp(e \varphi/T_\rme)$)
and that the densities of ions and electrons are equal at the internal (pre-sheath side) boundary of the sheath.
$n_0$ denotes the density at the boundary, and we set the basis of the electrostatic potential $\varphi$ at the boundary.
We also assume that the ion heat flux $\bm F$ obeys Fourier's law 
%($\bm F = - k \nabla T = - c_\rmv \chi n_0 \nabla T$; $k$ is the thermal conductivity,  $c_\rmv$ is the heat capacity at a constant volume, and $\chi = k / n_0 c_\rmv$ is the thermal diffusivity).
$\bm F = - c_\rmv \chi n_0 \nabla T$ with the heat capacity at a constant volume $c_\rmv$ and the thermal diffusivity $\chi$.
The evolution equation for the ion temperature $T$ is
\begin{gather}\label{eq:sagdif_temperature_0}
    c_\rmv n \left( \pdif{T}{t} + \bm{u} \cdot \nabla T \right) + p (\nabla \cdot \bm{u}) = - \nabla \cdot \bm F.
\end{gather}

We consider steady states on a one-dimensional system. 
We normalize the ion density $n$ by the density at the internal boundary $n_0$, 
the ion velocity $u$ by the ion sound speed without ion temperature $c_\rms=\sqrt{T_\rme/m}$ ($m$ is the ion mass),
the electrostatic potential $\varphi$ by the characteristic potential $T_\rme/e$,
the ion temperature $T$ by the electron temperature $T_\rme$,
the coordinate variable $x$ by the Debye length $\lambda_\rmD=\sqrt{\varepsilon_0 T_\rme/n_0 e^2}$,
and the thermal diffusivity $\chi$ by $\lambda_\rmD c_\rms$.
We obtain the mass conservation law $n u = M$ from the equation of continuity 
where $M$ is the ion inflow velocity.
The equation of motion (Bernoulli's law), with the expression of enthalpy for ideal gas $h = (c_\rmv + 1) T$, leads to
\begin{gather}\label{eq:sagdif_velocity} 
    u = M \left[ 1 - \frac{2\varphi}{M^2} - \frac{2 (c_\rmv + 1)}{M^2} (T - T_\mathrm{in}) \right]^{1/2},
\end{gather}
where $T_\mathrm{in}$ is the ion temperature at the internal boundary. 
The equation \eqref{eq:sagdif_temperature_0} leads to
\begin{gather}\label{eq:sagdif_temperature} 
    \chi \dif[2]{T}{x} -  M \dif{T}{x} - \frac{M}{c_\rmv} \dif{(\ln u)}{x} T = 0
\end{gather}
and the Poisson equation for the electrostatic potential $\varphi$ is
\begin{gather}\label{eq:sagdif_potential}
    \dif[2]{\varphi}{x} = \rme^\varphi - \frac{M}{u},
\end{gather}
where we substitute the expression \eqref{eq:sagdif_velocity} to $u$.

In the limit to the thermal diffusion without ion flows, 
we obtain a linear temperature profile ($\dif*[2]{T}{x} = 0$) from the equation \eqref{eq:sagdif_temperature}.
We solve the system \eqref{eq:sagdif_temperature}--\eqref{eq:sagdif_potential} and compare the results to the linear profile.
We note that 
in the opposite limit (without thermal diffusion), the equation \eqref{eq:sagdif_temperature} leads to the adiabatic relation $T n^{-1/c_\rmv} = \mathrm{const.}$ 
This relation enables us to write the equation \eqref{eq:sagdif_potential} by the Sagdeev potential.

%%%%%%%%%%%%%%%%%%%%%%%%%%%%%%%%%%%%%%%%%%%%%%%%%%%%%%%%%%%%%%%%
%% Numerical setting
%%%%%%%%%%%%%%%%%%%%%%%%%%%%%%%%%%%%%%%%%%%%%%%%%%%%%%%%%%%%%%%%

We consider the boundary-value problem of the equations \eqref{eq:sagdif_temperature}--\eqref{eq:sagdif_potential} on a one-dimensional space $[0,L]$, 
where $x=0$ is the internal edge and $x=L$ is the wall (here we put $L=10$).
At the wall boundary, we fix the temperature to a value $T(L) = T_\rmw$ (we put $T_\rmw = 0.1$). 
At the internal boundary, we control the heat flux $F_\mathrm{in}$, 
and the temperature $T_\mathrm{in}$ is free to vary. 
For the boundary value of the electrostatic potential, 
we assume the floating potential relation \cite{Stangeby2000}
\begin{gather}\label{eq:floating-potential}
    \varphi(L) = \varphi_\rmw := - \frac12 \ln \left( \frac{m}{2 \pi m_\rme} \right) + \ln M.
\end{gather}
Here we use $m \approx \SI{1.67e-27}{kg}$ (proton mass) and $m_\rme \approx \SI{9.11e-31}{kg}$ which lead to $\varphi_\rmw - \ln M \approx - 2.84$.

We use two control parameters: 
the boundary values of the heat flux and the ion Mach number defined by 
\begin{gather}\label{eq:Mach_temperature}
    \calM = \frac{M}{\sqrt{1 + \gamma T_\mathrm{in}}}
\end{gather}
($\gamma = 1+1/c_\rmv$ is the heat ratio).
We solve the equations \eqref{eq:sagdif_temperature}--\eqref{eq:sagdif_potential} iteratively 
with fixing the values of $F_\mathrm{in}$ and $\calM$. 
Solving the equation \eqref{eq:sagdif_temperature} changes $T_\mathrm{in}$. 
Hence we modify the values of $M$ and $\varphi_\rmw$ according to the equations \eqref{eq:Mach_temperature} and \eqref{eq:floating-potential}.

%%%%%%%%%%%%%%%%%%%%%%%%%%%%%%%%%%%%%%%%%%%%%%%%%%%%%%%%%%%%%%%%
%% Numerical results
%%%%%%%%%%%%%%%%%%%%%%%%%%%%%%%%%%%%%%%%%%%%%%%%%%%%%%%%%%%%%%%%

\begin{figure}[t]\centering
    \includegraphics[scale=.725]{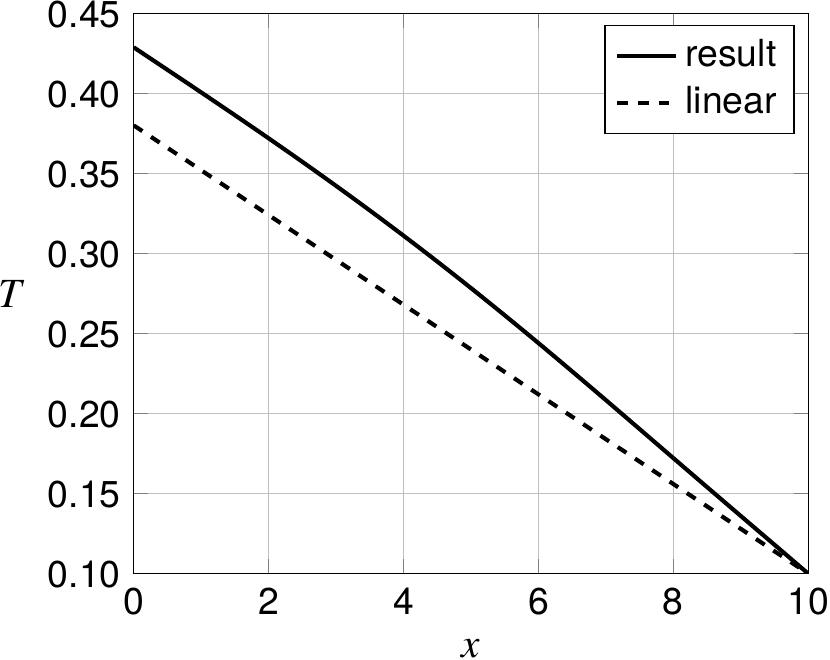}
    \caption{Temperature profile for $\calM=1$, $F_\mathrm{in} = 0.14$: contrast is larger than that of the linear profile.
    \label{fig:res_f-driven_M100}}
\end{figure}
\begin{figure}[t]\centering
    \includegraphics[scale=.725]{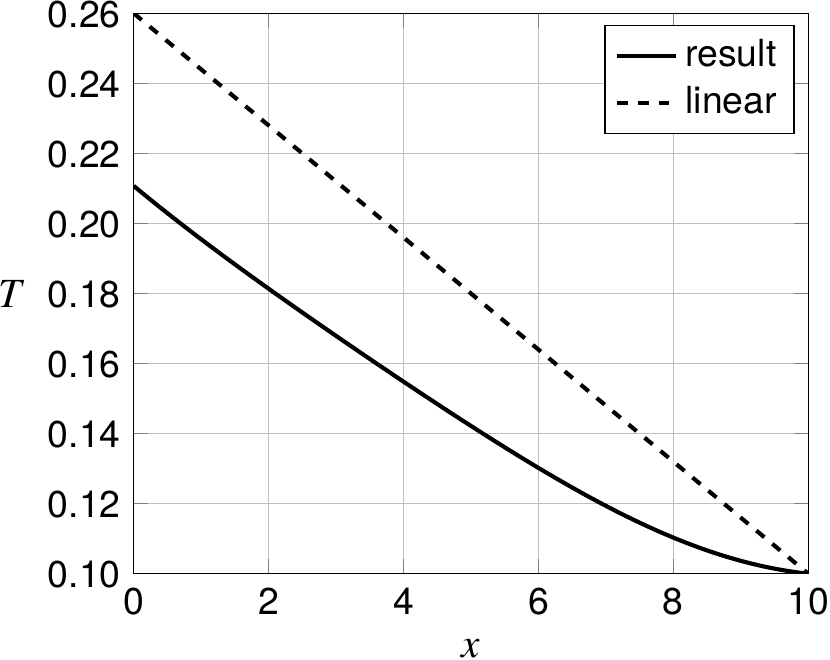}
    \caption{Temperature profile for $\calM=0.85$, $F_\mathrm{in} = 0.08$: contrast is smaller than that of the linear profile.
    \label{fig:res_f-driven_M085}}
\end{figure}

\begin{figure}[t]\centering
    \includegraphics[scale=.375]{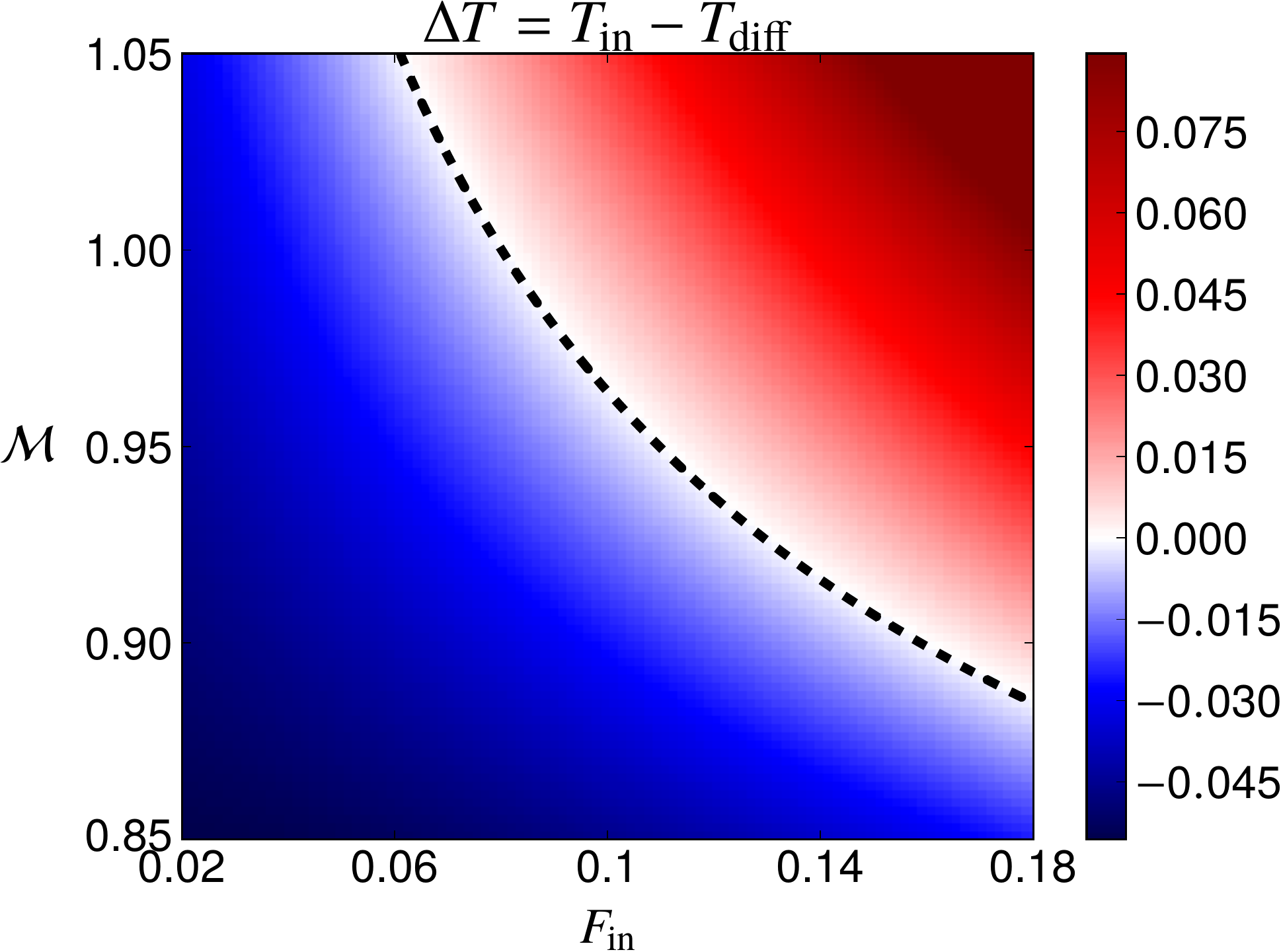}
    \caption{The difference between the internal boundary temperature and that of linear profile.
}
\label{fig:Dif_chi1e1}
\end{figure}

We present the results in Figures \ref{fig:res_f-driven_M100}--\ref{fig:Dif_chi1e1} 
(here we put $c_\rmv = 1/2$ and $\chi=10$).
\autoref{fig:res_f-driven_M100} shows the temperature profile 
with $\calM = 1$ and $F_\mathrm{in} = 0.14$.
We observe that the temperature contrast between boundaries is 
larger than that of the linear profile (thermal diffusion without ion flow).
\autoref{fig:res_f-driven_M085} shows the temperature profile 
with $\calM = 0.85$ and $F_\mathrm{in} = 0.08$.
In this case, we observe that the temperature contrast is 
smaller than that of the linear profile.
The transition from a smaller temperature contrast to larger one occurs depending on the boundary values of the heat flux $F_\mathrm{in}$ and the ion Mach number $\calM$. 
In \autoref{fig:Dif_chi1e1}, we present the differences between the temperature at the internal edge $T_\mathrm{in}$ and that of the linear profile $T_\mathrm{diff}$ ($\Delta T = T_\mathrm{in} - T_\mathrm{diff}$). 
The dashed line shows points where they coincide ($\Delta T = 0$). 
We observe $\Delta T < 0$ under the line (small $\calM$ and $F_\mathrm{in}$) and $\Delta T > 0$ over the line (large $\calM$ and $F_\mathrm{in}$).
We note that putting different values of the thermal diffusivity $\chi$ only moves the threshold line ($\Delta T = 0$) without changes of the property $\Delta T \gtrless 0$. 
Ions are accelerated to satisfy Bohm's criterion $\calM = 1$ in a pre-sheath region \cite{Stangeby2000,Riemann:sheath-review} 
(we note that a heat flux may modify the criterion \cite{TangGuo2016:Bohm}). 
%For simplicity, we assume that the ion inflow Mach number $\calM$ is fixed to the unity. 
Thus, we find that the change from $\Delta T < 0$ to $\Delta T > 0$ occurs when the heat flux exceeds a threshold value. 

The analyses of thermodynamical models \cite{YoshidaMahajan2008:MEP,KawazuraEPR1,KawazuraEPR2,YoshidaKawazuraEPR} elucidate that, 
in flux-driven systems, 
structures blocking heat transport cause a transition to larger temperature contrast state (maximization of EP)
and structures promoting heat transport cause a transition to smaller temperature contrast state (minimization of EP).
The observation obtained here indicates that the response of the sheath to heat transport 
changes from the latter type to the former type depending on the amount of the heat flux.

%%%%%%%%%%%%%%%%%%%%%%%%%%%%%%%%%%%%%%%%%%%%%%%%%%%%%%%%%%%%%%%%

This research was supported by JSPS KAKENHI Grant Number 15K13532.

%%%%%%%%%%%%%%%%%%%%%%%%%%%%%%%%%%%%%%%%%%%%%%%%%%%%%%%%%%%%%%%%

\bibliographystyle{pfr}
\bibliography{sheath_rapid}

\end{document}